# Efficient Multi-way Theta-Join Processing Using MapReduce

Xiaofei Zhang
HKUST
Hong Kong
zhangxf@cse.ust.hk

Lei Chen
HKUST
Hong Kong
leichen@cse.ust.hk

Min Wang
HP Labs China
Beijing, China
min.wang6@hp.com

## ABSTRACT

Multi-way Theta-join queries are powerful in describing complex relations and therefore widely employed in real practices. However, existing solutions from traditional distributed and parallel databases for multi-way Theta-join queries cannot be easily extended to fit a shared-nothing distributed computing paradigm, which is proven to be able to support OLAP applications over immense data volumes. In this work, we study the problem of efficient processing of multi-way Theta-join queries using MapReduce from a cost-effective perspective. Although there have been some works using the (*key,value*) pair-based programming model to support join operations, efficient processing of multi-way Theta-join queries has never been fully explored. The substantial challenge lies in, given a number of processing units (that can run Map or Reduce tasks), mapping a multi-way Theta-join query to a number of MapReduce jobs and having them executed in a well scheduled sequence, such that the total processing time span is minimized. Our solution mainly includes two parts: 1) cost metrics for both single MapReduce job and a number of MapReduce jobs executed in a certain order; 2) the efficient execution of a chain-typed Theta-join with only one MapReduce job. Comparing with the query evaluation strategy proposed in [23] and the widely adopted Pig Latin and Hive SQL solutions, our method achieves significant improvement of the join processing efficiency.

## 1. INTRODUCTION

Data analytical queries in real practices commonly involve multi-way join operations. The operators involved in a multi-way join query are more than just Equi-join. Instead, the join condition can be defined as a binary function $\theta$ that belongs to $\{<, \leq, =, \geq, >, <>\}$, as known as Theta-join. Compared with Equi-join, it is more general and expressive in relation description and surprisingly handy in data analytic queries. Thus, efficient processing of multi-way Theta-join queries plays a critical role in the system performance. In fact, evaluating multi-way Theta-joins has always been a challenging problem along with the development of database technology. Early works, like [8][26][22] and etc., have elaborated the complexity of the problem and presented their evaluation strategies. However, their solutions do not scale to process the multi-way Theta-joins over the data of tremendous volumes. For instance, as reported from Facebook [5] and Google [11], the underlying data volume is of hundreds of tera-bytes or even peta-bytes. In such scenarios, solutions from the traditional distributed or parallel databases are infeasible due to unsatisfactory scalability and poor fault tolerance.

On the contrary, (*key,value*)-based MapReduce programming model substantially guarantees great scalability and strong fault tolerance property. It has emerged as the most popular processing paradigm in a shared-nothing computing environment. Recently, devoting research efforts towards efficient and effective analytic processing over immense data have been made within the MapReduce framework. Currently, the database community mainly focuses on two issues. First, the transformation from certain relational algebra operator, like similarity join, to its (*key,value*)-based parallel implementation. Second, the tuning or re-design of the transformation function such that the MapReduce job is executed more efficiently in terms of less time cost or computing resources consumption. Although various relational operators, like pair-wise Theta-join, fuzzy join, aggregation operators and etc., are evaluated and implemented using MapReduce, there is little effort exploring the efficient processing of multi-way join queries, especially more general computation namely Theta-join, using MapReduce. The reason is that, the problem involves more than just a *relational operator*→(*key,value*) pair transformation and the tuning, there are other critical issues needed to be addressed: 1) How many MapReduce jobs should we employ to evaluate the query? 2) What is each MapReduce job responsible for? 3) How should multiple MapReduce jobs be scheduled?

To address the problem, there are two challenging issues needed to be resolved. Firstly, the number of available computing units is in fact limited, which is often neglected when mapping a task to a set of MapReduce jobs. Although the *pay-as-you-go* policy of Cloud computing platform could promise as many computing resources as required, however, once a computing environment is established, the allowed maximum number of concurrent Map and Reduce tasks is fixed according to the system configuration. Even taken the *auto scaling* feature of Amazon EC2 platform [18] into consideration, the maximum number of involved computing units are pre-determined by the user-defined profiles. There-fore, with the user specified Reduce task number, a multi-way Theta-join query is processed with only limited number of available computing units.

The second challenge is that, the decomposition of a multi-way Theta-join query into a number of MapReduce tasks is non-trivial. Work [28] targets at the multi-way Equi-join processing. It decomposes a query into several MapReduce jobs and schedules the execution based on a specific cost model. However, it only considers the pair-wise join as the basic scheduling unit. In other words, it follows the traditional multi-way join processing methodology, which evaluates the query with a sequence of pair-wise joins. This methodology excludes the possible optimization opportunity to evaluate a multi-way join in one MapReduce job. Our observation is that, under certain conditions, evaluating a multi-way join with one MapReduce job is much more efficient than with a sequence of MapReduce jobs conducting pair-wise joins. Work [23] reports the same observation. One dominating reason is that, the I/O costs of intermediate results generated by multiple MapReduce jobs may become unacceptable overheads. Work [2] presents the solution of evaluating a multi-way join in one MapReduce job, which only works for the Equi-join case. Since the Theta-join cannot be answered by simply making the join attribute the partition key, thus, the solution proposed in [2] cannot be extended to solve the case of multi-way Theta-joins. Work [25] demonstrates effective pair-wise Theta-join processing using MapReduce by partitioning a two dimensional result space formed by the cross-product of two relations. For the case of multi-way join, the result space is a hyper-cube, whose dimensionality is the number of the relations involved in the query. Unfortunately, work [25] does not explore how to extend their solution to handle the partition in high dimensions. Moreover, the question about whether we should evaluate a complex query with a single MapReduce job or several MapReduce jobs, is not clear yet. Therefore, there is no straightforward solution to combine the techniques in existing literatures to evaluate a multi-way Theta-join query.

Meanwhile, assume a set of MapReduce jobs are generated for the query evaluation. Then given a limited number of processing units, it remains a challenge to schedule the execution of MapReduce jobs, such that the query can be answered with the minimum time span. These jobs may have dependency relationships and inter-competition for resource consumptions during the concurrent execution. Currently, the MapReduce framework requires the number of Reduce tasks as a user specified input. Thus, after decomposing a multi-way Theta-join query into a number of MapReduce jobs, one challenging issue is how to specify each job a proper Reduce task number, such that the overall scheduling achieves the minimum execution time span.

Specifically, the problem that we are working on is: given a number of processing units (that can run Map or Reduce tasks), mapping a multi-way Theta-join to a number of MapReduce jobs and having them executed in a well scheduled order, such that the total processing time span is minimized. Our solution to this challenging problem includes two core techniques. The first one is, given a multi-way Theta-join query, we examine all the possible decomposition plans and estimate the minimum execution time cost for each plan. Especially, we figure out the rules to properly decompose the original multi-way Theta-join query and study the most efficient solution to evaluate multiple join condition functions using one MapReduce job. The second technique is that, given a limited number of computing units and a pool of possible MapReduce jobs to evaluate the query, we design a novel solution to select jobs to effectively evaluate the query as fast as possible. To evaluate the cost, we develop an I/O and network aware cost model to describe the behavior of a MapReduce job.

To the best of our knowledge, this is the first work exploring the multi-way Theta-joins evaluation using MapReduce. Our main contributions are listed as follows:

- We establish the rules to decompose a multi-way join query. Under our proposed cost model, we can figure out whether a multi-way join query should be evaluated with multiple MapReduce jobs or a single MapReduce job.
- We develop a resource aware (*key,value*) pair distribution method to evaluate the chain-typed multi-way Theta-join query with one MapReduce job, which guarantees minimized volume of data copying over the network, as well as evenly distributed workload among Reduce tasks.
- We validate our cost model and the solution for multi-way Theta-join queries with extensive experiments.

The rest of the paper is organized as follows. In Section 2, we briefly review the MapReduce computing paradigm and elaborate the application scenario for multi-way Theta-joins. We formally define our problem in Section 3 and present our cost model in section 4. We take Section 5 to explain our query evaluation strategies in details. We validate our solution in Section 6 with extensive experiments on both real and synthetic data sets. We summarize and compare the most recent closely related work in Section 7 and conclude our work in Section 8.

## 2. PRELIMINARIES

In this section we briefly present the MapReduce programming model and how it has been applied to evaluate join queries. More importantly, we elaborate the difficulties and limitations of current solutions to solve the multi-way Theta-joins with a concrete example.

### 2.1 MapReduce & Join Processing

MapReduce provides a simple parallel programming model for data-intensive applications in a shared-nothing environment [12]. It was originally developed for indexing crawled websites and OLAP applications. Generally, a *Master* node invokes Map tasks on computing nodes that possess the input data, which guarantees the locality of computation. Map tasks transform the input (*key,value*) pair $(k^1,v^1)$ to $n$ new pairs: $(k_1^2,v_1^2)$, $(k_2^2,v_2^2)$, ..., $(k_n^2,v_n^2)$. The output of Map tasks are then partitioned by the default hashing to different Reduce tasks according to $k_i^2$. Once the Reduce tasks receive (*key,value*) pairs grouped by $k_i^2$, they perform the user specified computation on all the values of each *key*, and write results back to the storage.

Obviously, this (*key,value*)-based programming model implies a natural implementation of Equi-join. By making the join attribute the *key*, records that can be joined together are sent to the same Reduce task. Even for the similarity join case [27], as long as the similarity metric is defined, each data record is assigned with a *key* set $\mathcal{K} = \{k_i, ..., k_j\}$, and the intersection of similar data records' *key* sets is never empty. Thus, through such a mapping, it guarantees that similar data records will be sent to at least one common Reduce task.

In fact, this *key set* method can be applied to any type of join operator. However, to ensure that joinable data records are always assigned to overlapping *key* sets, the cardinality of a data record's $\mathcal{K}$ can be very large. In the worst case, it is the total number of Reduce tasks. Since the cardinality of a record's $\mathcal{K}$ implies the number of times this record being duplicated among Reduce tasks, the larger the value is, the more computing overheads in terms of I/O and CPU consumption will be introduced. Therefore, the essential optimization goal is to find "the optimal" assignment of $\mathcal{K}$ to each data record, such that the join query can be evaluated with minimized data transmission over the network.

Another common concern about the MapReduce programming model is its poor immunity to *key* skews. If (*key*,*value*) pairs are highly unevenly distributed among Reduce tasks, the system throughput can degrade significantly. Unfortunately, this could be a common scenario in join operations. If there exist "popular" join attribute values, or the join condition is an inequality, some data records can be joined with huge number of data records from other relations, which implies significant *key* skew among the Reduce tasks. Moreover, the fault tolerance property of the MapReduce programming model is guaranteed on the cost of saving all the intermediate results. Thus, the overhead of disk I/O dominates the time efficiency of iterative MapReduce jobs. The same observation has been made in [28].

In summary, to efficiently process join operations using MapReduce is non-trivial. Especially when it comes to multi-way join processing, selecting proper MapReduce jobs and deciding a proper $\mathcal{K}$ for each data record make the problem more challenging.

## 2.2 Multi-way Theta-Join

Theta-join is the join operation that takes inequality conditions of join attributes' values into consideration, namely the join condition function $\theta \in \{<, >, =, <>, \leq, \geq\}$. Multi-way Theta-join is a powerful analytic tool to elaborate complex data correlations. Consider the following application scenario:

"Assume we have $n$ cities, $\{c_1, c_2, ..., c_n\}$, and all the flights information $FI_{i,j}$ between any two cities $c_i$ and $c_j$. Given a sequence of cities $< c_s, ..., c_t >$, and the stay-over time length which must fall in the interval $L_i = [l_1, l_2]$ at each city $c_i$, find out all the possible travel plans."

This is a practical query that could help travelers plan their trips. For illustration purpose, we simply assume $FI_{i,j}$ is a table containing flight No., departure time (*dt*) and arrival time (*at*). Then the above request can be easily answered with a multi-way Theta-join operation over $FI_{s,s+1}$, ..., $FI_{t-1,t}$, by specifying the time interval between two successive flights falling into the particular city's stay-over interval requirement. For example, the $\theta$ function between $FI_{s,s+1}$ and $FI_{s+1,s+2}$ is $FI_{s,s+1}.at + L_{s+1}.l_1 < FI_{s+1,s+2}.dt < FI_{s,s+1}.at + L_{s+1}.l_2$.

To evaluate such queries, a straightforward method is to iteratively conduct pair-wise Theta-join. However, this evaluation strategy might exclude some more efficient evaluation plans. For instance, instead of using pair-wise joins, we can evaluate multiple join conditions in one task. Therefore, less MapReduce jobs are needed, which implies less computation overheads in terms of the disk I/O of intermediate results.

## 3. PROBLEM DEFINITION

In this work, we mainly focus on the efficient processing of multi-way Theta-joins using MapReduce. Our solution targets on the MapReduce job identification and scheduling. In other words, we work on the rules to properly decompose the query processing into several MapReduce jobs and have them executed in a well scheduled fashion, such that the minimum evaluation time span is achieved. In this section, we shall first present the terminologies that we use in this paper, and then give the formal definition of the problem. We show that the problem of finding the optimal query evaluation plan is NP hard.

### 3.1 Terminology and Statement

For the ease of presentation, in the rest of the paper we use the notation of "N-join" query to denote a multi-way Theta-join query. We use MRJ to denote a MapReduce job.

Consider a N-join query $\mathcal{Q}$ defined over $m$ relations $\mathcal{R}_1, ..., \mathcal{R}_m$ and $n$ specified join conditions $\theta_1, ..., \theta_n$. As adopted in many other works, like in [28], we can present $\mathcal{Q}$ as a graph, namely a join graph. For completeness, we define a join graph $\mathcal{G}_J$ as follows:

**Definition 1** *A join graph $\mathcal{G}_J = \langle V, E, L \rangle$ is a connected graph with edge labels, where $V = \{v | v \in \{\mathcal{R}_1, ..., \mathcal{R}_m\}\}$, $E = \{e | e = (v_i, v_j) \iff \exists \theta, \mathcal{R}_i \bowtie_\theta \mathcal{R}_j \in \mathcal{Q}\}$, $L = \{l | l(e_i) = \theta_i\}$.*

Intuitively, $\mathcal{G}_J$ is generated by making every relation in $\mathcal{Q}$ a vertex and connecting two vertices if there is a join operator between them. The edge is labeled with the corresponding join function $\theta$. To evaluate $\mathcal{Q}$, every $\theta$ function, i.e., every edge from $\mathcal{G}_J$, needs to be evaluated. However, to evaluate all the edges in $\mathcal{G}_J$, there are exponential number of plans since any arbitrary number of connecting edges can be evaluated in one MRJ. We propose a join-path graph to cover all the possibilities. For the purpose of clear illustration, we define a no-edge-repeating path between two vertices of $\mathcal{G}_J$ in the first place.

**Definition 2** *A no-edge-repeating path $p$ between two vertices $v_i$ and $v_j$ in $\mathcal{G}_J$ is a traversing sequence of connecting edges $\langle e_i, ..., e_j \rangle$ between $v_i$ and $v_j$ in $\mathcal{G}_J$, in which no edge appears more than once.*

**Definition 3** *A join-path graph $\mathcal{G}_{JP} = \langle V, E', L', W, S \rangle$ is a complete weighted graph with edge labels, where each edge is associated with a weight and scheduling information. Specifically, $V = \{v | v \in \{\mathcal{R}_1, ..., \mathcal{R}_m\}\}$, $E' = \{e' | e' = (v_i, v_j) \text{ represents a unique no-edge-repeating path } p \text{ between } v_i \text{ and } v_j \text{ in } \mathcal{G}_J\}$, $L' = \{l' | l'(e') = l'(v_i, v_j) = \bigcup l(e), e \in p \text{ between } v_i \text{ and } v_j\}$, $W = \{w | w(e') \text{ is the minimal cost to evaluate } e'\}$, $S = \{s | s(e') \text{ is the scheduling to evaluate } e' \text{ at the cost of } w(e')\}$.*

In the definition, the scheduling information on the edge refers to some user specified parameter to run a MRJ, such that this job is expected to be accomplished as fast as possible. In this work, we consider the number of Reduce tasks assigned to a MRJ as the scheduling parameter, denoted as $RN(\text{MRJ})$, as it is the only parameter that users need to specify in their programs. The reason we take this parameter into consideration is based on two observations from extensive experiments: 1) It is not guaranteed that the more computing units involved in Reduce tasks, the sooner a MRJ job is accomplished; 2) Given limited computing units, there is resource competition among multiple MRJs.

Intuitively, we enumerate all the possible join combinations in $\mathcal{G}_{JP}$. Note that in the context of join processing, $\mathcal{R}_i \bowtie \mathcal{R}_k \bowtie \mathcal{R}_j$ is the same with $\mathcal{R}_j \bowtie \mathcal{R}_k \bowtie \mathcal{R}_i$, therefore, $\mathcal{G}_{JP}$ is an undirected graph. We elaborate Definition 3.3 with the following example. Given a join graph $\mathcal{G}_J$, shown on the left in Fig.1, a corresponding join-path graph $\mathcal{G}_{JP}$ is generated, which is presented in an adjacent matrix format on the right. The numbers enclosed in bracelets are the involved $\theta$ functions on a path. For instance, in the cell corresponding to $R_1$ and $R_2$, $\{3, 4, 6, 5, 2\}$ indicates a no-edge-repeating path $\{\theta_3, \theta_4, \theta_6, \theta_5, \theta_2\}$ between $R_1$ and $R_2$. For this particular example, notice that for every node there exists a closed traversing path (or circuit) which covers all the edges exactly once, namely the "Eulerian Circuit". We use $\mathcal{E}(\mathcal{G}_{JP})$ to denote a "Eulerian Circuit" of $\mathcal{G}_{JP}$ in the figure. Since we only care what edges are involved in a path, any $\mathcal{E}(\mathcal{G}_{JP})$ would be sufficient. Notice that in the figure, edge weights and scheduling information are not presented. As a matter of fact, these information are incrementally computed during the generation of $\mathcal{G}_{JP}$, which will be illustrated in the later Section.

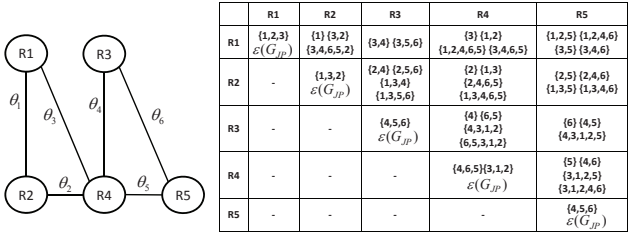

**Figure 1: Example join graph $\mathcal{G}_J$ and its corresponding join-path graph $\mathcal{G}_{JP}$, presented in an adjacent matrix**

According to the definition of $\mathcal{G}_{JP}$, any edge $e'$ in $\mathcal{G}_{JP}$ is a collection of connecting edges in $\mathcal{G}_J$. Thus, $e'$ in fact implies a subgraph of $\mathcal{G}_J$. As we use one MRJ to evaluate $e'$, denoted as $MRJ(e')$, $\mathcal{G}_{JP}$'s edge set represents all the possible MRJs that can be employed to evaluate the original query $\mathcal{Q}$. Let $\mathcal{T}$ denote a set of MRJs that are selected from $\mathcal{G}_{JP}$'s edge set. Intuitively, if the MRJs in $\mathcal{T}$ cover all the join conditions of the original query, we can answer the query by executing all these MRJs. Formally, we define that $\mathcal{T}$ is "sufficient" as follows:

**Definition 4** $\mathcal{T}$, a collection of MRJs, is sufficient to evaluate $\mathcal{Q}$ iff $\bigcup e'_i = \mathcal{G}_J.E$, where $MRJ(e'_i) \in \mathcal{T}$,

Since it is trivial to check whether $\mathcal{T}$ is sufficient, for the rest of this work, we only consider the case that $\mathcal{T}$ is sufficient. Thus, given $\mathcal{T}$, we define its execution plan $\mathcal{P}$ as a specific execution sequence of MRJs, which minimizes the time span of using $\mathcal{T}$ to evaluate the original query $\mathcal{Q}$. Formally, we can define our problem as follows:

**Problem Definition:** Given a N-join query $\mathcal{Q}$ and $k_P$ processing units, a join-path graph $\mathcal{G}_{JP}$ according to $\mathcal{Q}$'s join graph $\mathcal{G}_J$ is built. We want to select a collection of edges from $\mathcal{G}_{JP}$ that correspondingly form a set of MRJs, denoted as $\mathcal{T}_{opt}$, such that there exists an execution plan $\mathcal{P}$ of $\mathcal{T}_{opt}$ which minimizes the query evaluation time.

Obviously, there are many different choices of $\mathcal{T}$ to evaluate $\mathcal{Q}$. Moreover, given $\mathcal{T}$ and limited processing units, different execution plans yield different evaluation time spans. In fact, the determination of $\mathcal{P}$ is non-trivial, we give the detailed analysis of the hardness of our problem in the next subsection. As we shall elaborate later, given $\mathcal{T}$ and $k_P$ available processing units, we adopt an approximation method to determine $\mathcal{P}$ in linear time.

## 3.2 Problem Hardness

According to the problem definition, we need two steps to find $\mathcal{T}_{opt}$: 1) generate $\mathcal{G}_{JP}$ from $\mathcal{G}_J$; 2) select MRJs for $\mathcal{T}_{opt}$. Neither one of these two steps is easy to solve.

For the first step, to construct $\mathcal{G}_{JP}$, we need to enumerate all the no-edge-repeating paths between any pair of vertices in $\mathcal{G}_J$. Assume $\mathcal{G}_J$ has the "Eulerian trail"[16], which is a way to traverse the graph with every edge be visited exactly once, then for any pair of vertices $v_i$ and $v_j$, any different no-edge-repeating path between them is a "sub-path" of an Eulerian trail. If we know all the no-edge-repeating paths between any pair of vertices, we can enumerate all the Eulerian trails in polynomial time. Therefore, the complexity of constructing $\mathcal{G}_{JP}$ is at least as hard as enumerating all the Eulerian trails of a given graph, which is known to be #$\mathcal{P}$-complete [6]. Moreover, we find that even $\mathcal{G}_J$ does not have an Eulerian trail, the problem complexity is not reduced at all, as we elaborate in the proof of the following theorem.

**Theorem 1** Generating $\mathcal{G}_{JP}$ from a given $\mathcal{G}_J$ is a #$\mathcal{P}$ complete problem.

PROOF. If $\mathcal{G}_J$ has the Eulerian trail, constructing $\mathcal{G}_{JP}$ is #$\mathcal{P}$-complete (see the discussion above).

On the contrary, if $\mathcal{G}_J$ does not have the Eulerian trail, it implies that there are $r$ vertices having odd degrees, where $r > 2$. Now consider that we add one virtual vertex and connecting it with $r$-1 vertices of odd degrees. Now the graph must have an Eulerian trail. If we can easily construct the join-path graph of the new graph, the original graph's $\mathcal{G}_{JP}$ can be computed in polynomial time. We elaborate with the following example, as shown in Fig.2. Assume $v_s$ is added to the original $\mathcal{G}_J$, then by computing the join-path graph of the new graph, we know all the no-edge-repeating paths between $v_i$ and $v_j$. Then, a no-edge-repeating path between $v_i$ and $v_j$ cannot exist if it has $v_s$ involved. By simply removing all the enumerated paths that go through $v_s$, we can obtain the $\mathcal{G}_{JP}$ of the original $\mathcal{G}_J$. Thus, the dominating cost of constructing $\mathcal{G}_{JP}$ is still the enumeration of all Eulerian trails. Therefore, this problem is #$\mathcal{P}$-complete. □

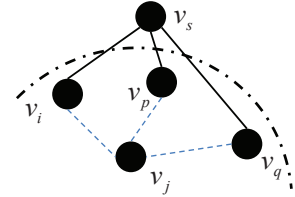

**Figure 2: Adding virtual vertex $v_s$ to $\mathcal{G}_J$**

Although it is difficult to compute the exact $\mathcal{G}_{JP}$, we find that a subgraph of $\mathcal{G}_{JP}$, which contains all the vertices and denoted as $\mathcal{G}'_{JP}$, could be sufficient to guarantee the optimal query evaluation efficiency. We take the following principle into the consideration. Given the same number of processing units, if it takes longer time to evaluate $\mathcal{R}_i \bowtie \mathcal{R}_j \bowtie \mathcal{R}_k$ with one MRJ compared to the total time cost of evaluating $\mathcal{R}_i \bowtie \mathcal{R}_j$ and $\mathcal{R}_j \bowtie \mathcal{R}_k$ separately and merging the results, we do not take $\mathcal{R}_i \bowtie \mathcal{R}_j \bowtie \mathcal{R}_k \bowtie \mathcal{R}_s$ into consideration. By following this principle, we can avoid enumerating all the possible no-edge-repeating paths between any pair of vertices. As a matter of fact, we can obtain such a sufficient $\mathcal{G}'_{\text{JP}}$ in polynomial time.

The second step of our solution is to select the $\mathcal{T}_{opt}$. Assume the $\mathcal{G}'_{\text{JP}}$ computed from the first step provides a collection of edges, accordingly, we have a collection of MRJ candidates to evaluate the query. Although each edge in $\mathcal{G}_{\text{JP}}$ is associated with a weight denoting the minimum time cost to evaluate all the join conditions contained in this edge, it is just an estimated time span on the condition that there are enough processing units. However, when a $\mathcal{T}$ is chosen, and the number of processing units is limited, the time cost of using $\mathcal{T}$ to answer $\mathcal{Q}$ need to be re-estimated. Assume we can find the time cost estimation of $\mathcal{T}$, denoted as $C(\mathcal{T})$, then the problem is to find such an optimal $\mathcal{T}_{opt}$ from all possible $\mathcal{T}$s, which has the minimum time cost. Apparently, this is a variance of the classic set cover problem, which is known to be NP hard [10]. Therefore, there are many heuristics and approximation algorithms can be adopted to solve the selection problem.

As clearly indicated in the problem definition, the solution lies in the construction of $\mathcal{G}'_{\text{JP}}$ and smartly select $\mathcal{T}$ based on the cost estimation of a group of MRJs. Therefore, for the rest of the paper, we shall first elaborate our cost models for a single MRJ and a group of MRJs. Then we present our detailed solution for the N-join query evaluation.

## 4. COST MODEL

To highlight our observations on how much the overlapping of computation and network cost would affect the execution of a MRJ, in this section we present a generalized analytical study on the execution time of both a single MRJ and a group of MRJs. In the context of $\mathcal{G}_{\text{JP}}$ construction and $\mathcal{T}$ selection, we study the estimation of $w(e')$, where $e' \in \mathcal{G}_{\text{JP}}.E$, and $C(\mathcal{T})$, which is the time cost to evaluate $\mathcal{T}$.

### 4.1 Estimating $w(e')$: Model for Single MRJ

Since our target is to find an optimal join plan, we only consider the processing cost of join operations with MRJs. Generally, most of the CPU time for join processing is spent on simple comparison and counting, thus, system I/O cost dominates the total execution time. For MapReduce jobs, heavy cost on large scale sequential disk scan and frequent I/O of intermediate results dominate the execution time. Therefore, we shall build a model for a MRJ's execution time based on the analysis of I/O and network cost.

General MapReduce computing framework involves three phases of data processing: Map, Reduce and the data copying from Map tasks to Reduce tasks, as shown in Fig.3. In the figure, each "M" stands for a Map task; each "CP" stands for one phase of Map output copying over network, and each "R" stands for a Reduce task. Since each Map task is based on a data block, we assume that the unit processing cost for each Map task is $t_M$. Moreover, since the entire input data may not be loaded into the system memory within one round [12] [3], we assume these Map tasks are performed round by round (we have the same observation in practice). However, the size of Reduce task is subjected to the $(key, value)$ distribution. As shown in Fig.3, the make span of a MRJ is dominated by the most time consuming Reduce task. Therefore, we only consider the Reduce task with the largest volume of inputs in the following analysis. Assume the total input size of a MRJ is $S_I$, the total intermediate data copied from Map to Reduce is of size $S_{\text{CP}}$, the number of Map tasks and Reduce tasks are $m$ and $n$, respectively. In addition, as a general assumption, $S_I$ is considered to be evenly partitioned among $m$ Map tasks [24]. Let $J_M$, $J_R$ and $J_{\text{CP}}$ denote the total time cost of three phases respectively, $T$ be the total execution time of a MRJ. Then $T \leq J_M + J_{\text{CP}} + J_R$ holds due to the overlapping between $J_M$ and $J_{\text{CP}}$.

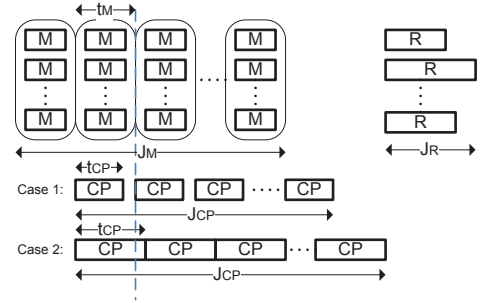

Figure 3: MapReduce workflow

For each Map task, it performs disk I/O and data processing. Since disk I/O is the dominant cost, therefore, we can estimate the time cost for single Map task based on disk I/O. Disk I/O contains two parts, one is sequential reading, the other is data spilling. Then the time cost for single Map Task $t_M$ is

$$t_M = (C_1 + p \times \alpha) \times \frac{S_I}{m} \quad (1)$$

where $C_1$ is a constant factor regarding disk I/O capability, $p$ is a random variable denoting the cost of spilling intermediate data. For a given system configuration, $p$ subjects to the intermediate data size; it increases while spilled data size grows. $\alpha$ denotes the output ratio of a Map task, which is query specific and can be computed with the selectivity estimation. Assume $m'$ is the current number of Map tasks running in parallel in the system, then $J_M$ can be computed as follows

$$J_M = t_M \times \frac{m}{m'} \quad (2)$$

For $J_{\text{CP}}$, let $t_{CP}$ be the time cost for copying the output of single Map task to $n$ Reduce tasks, it includes the cost of data copying over network as well as overhead of serving network protocols. $t_{\text{CP}}$ is calculated with the following formula,

$$t_{\text{CP}} = C_2 \times \frac{\alpha \times S_I}{n \times m} + q \times n \quad (3)$$

where $C_2$ is a constant number denoting the efficiency of data copying over network, $q$ is a random variable which represents the cost of a Map task serving $n$ connections from $n$ Reduce tasks. Intuitively, there is a rapid growth of $q$ while $n$ gets larger. Thus, $J_{\text{CP}}$ can be computed as follows:

$$J_{\text{CP}} = \frac{m}{m'} \times t_{\text{CP}} \quad (4)$$

For $J_R$, intuitively it is dominated by the Reduce task which has the biggest size of input. We assume that the key distribution in the input file is random; thus let $S_r^i$ denote the input size of Reduce task $i$, then according to the *Central Limit Theorem*[20], we can assume for $i = 1, ..., n$, $S_r^i$ follows a normal distribution $N \sim (\mu, \sigma)$, where $\mu$ is determined by $\alpha \times S_I$ and $\sigma$ subjects to data set properties, which can be learned from history query logs. Thus, by employing the rule of "*three sigmas*"[20], we make $S_r^* = \alpha \times S_I \times n^{-1} + 3\sigma$ the biggest input size to a Reduce task, then

$$J_R = (p + \beta \times C_1) \times S_r^* \quad (5)$$

where $\beta$ is a query dependent variable denoting output ratio, which could be pre-computed based on the selectivity estimation.

Thus, the execution time $T$ of a MRJ is:

$$T = \begin{cases} J_M + t_{CP} + J_R & \text{if } t_M \geq t_{CP} \\ t_M + J_{CP} + J_R & \text{if } t_M \leq t_{CP} \end{cases} \quad (6)$$

In our cost model, parameters $C_1$, $C_2$, $p$ and $q$ are system dependent and need to be derived from observations on the execution of real jobs, which are elaborated in the experiments section. This model favors MRJs that have I/O cost dominate the execution time. Experiments show that our method can produce a reasonable approximation of the MRJ running time in real practice.

### 4.2 Estimating $C(\mathcal{T})$: Model for A Group of MRJs

There have been some works exploring the optimization opportunity among multiple MRJs running in parallel, like [23] [24] and [28], by defining multiple types of correlations among MRJs. For instance, [23] defines "input correlation", "transit correlation" and "job flow correlation", targeting at the shared input scan and intermediate data partition. In fact, their techniques can be directly plugged into our solution framework. Compared to these techniques, the significant difference of our study on the execution model of a set of MRJs is that our work takes the number of available processing units into consideration. Therefore, the optimization problem we study here is orthogonal with the techniques proposed in existing literatures that we mentioned above.

Given $\mathcal{T}$ and $k_P$ processing units, we concern about the execution plan $\mathcal{P}$ that guarantees the minimum task execution time span. However, the determination of $\mathcal{P}$ is usually subjected to $k_P$. For instance, consider the $\mathcal{T}$ given in Fig.4. MRJ($e'_i$), MRJ($e'_j$) and MRJ($e'_k$) can be accomplished in 5, 7, 9 time units if 4, 4, 8 Reduce tasks are assigned to them respectively. Thus, if there are over 16 available processing units, these three MRJs can be scheduled to run in parallel and have no computing resource competition. On the contrary, if there are not enough processing units, parallel execution of multiple MRJs can lead to very poor time efficiency. It is exactly the classic problem of scheduling independent malleable parallel tasks over bounded parallel processors, which is NP hard [19]. In this work, we adopt the methodology presented in [19]. The method guarantees that for any given $\epsilon > 0$, it takes linear time (in terms of $|\mathcal{T}|$, $k_P$ and $\epsilon^{-1}$) to compute a scheduling that promises the evaluation time to be at most $(1+\epsilon)$ times the optimal one.

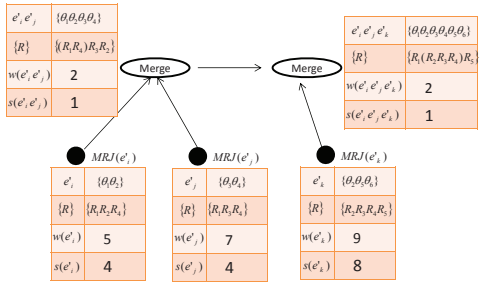

**Figure 4: One execution plan of $\mathcal{T}=\{e'_i, e'_j, e'_k\}$**

Moreover, to evaluate $\mathcal{Q}$ with $\mathcal{T}$, not only the MRJs in $\mathcal{T}$ must be executed, a merge step is needed to generate the final results. Intuitively, if two MRJs share some common input relation, their output can be merged using the common relation as the key. For instance, Fig.4 presents one possible execution plan of MRJ($e'_i$), MRJ($e'_j$) and MRJ($e'_k$). Assume there are over 16 available processing units, then we execute all three jobs in parallel. Since MRJ($e'_i$) and MRJ($e'_j$) share the same input $\mathcal{R}_1$ and $\mathcal{R}_4$. Therefore, the output of MRJ($e'_i$) and MRJ($e'_j$) can be merged using the primary keys from both $\mathcal{R}_1$ and $\mathcal{R}_4$. Later on, the output of this step can be further merged with the output of MRJ($e'_k$). The total execution time is 9+2=11 time units. In the figure, we enclose the merge key with bracket. Note that such a merge operation only has output keys or data IDs involved, therefore, it can be done very efficiently.

## 5. JOIN ALGORITHM

As discussed in Section 3, the key issues of our solution lie in constructing $\mathcal{G}'_{JP}$ and selecting $\mathcal{T}$. In section 4, we present an analytical study of the execution schedules of a single MRJ and multiple MRJs. However, we have not yet solve the problem of how to compute a multi-way Theta-join in one MRJ. Therefore, in this section, we first present our solution to the multi-way Theta-join processing with one MRJ. Then, we elaborate the construction of $\mathcal{G}'_{JP}$ and the selection of $\mathcal{T}$.

### 5.1 Multi-way Theta-join Processing with Single MRJ

As discussed in Section 2, different from Equi-join, we cannot use the join attribute as the hash *key* to answer Theta-join in the MapReduce computing framework. Work [25] for the first time explores the way to adopt MapReduce to answer a Theta-join query. Essentially, it partitions the cross-product result space with rectangle regions of bounded size, which guarantees the output correctness and the workload balance among Reduce tasks. However, their partition method does not have a straightforward extension to solve a multi-way Theta-join query. Inspired from work [25], we believe that it is a feasible solution to conceptually make the cross-product of multiple relations as the starting point and figure out a better partition strategy.

Based on our problem definition, all the possible MRJ candidates for $\mathcal{T}$ is a no-edge-repeating path in the join graph $\mathcal{G}_J$. Thus, we only consider the case of chain joins. Given a chain Theta-join query with $m$ different relations involved, we want to derive a ($key,value$)-based solution that guarantees the minimum execution time span. Let $\mathcal{S}$ denote the hyper-cube that comprises the cross-product of all $m$ relations. Let $f$ denote a space partition function that maps $\mathcal{S}$ to a set of disjoint components whose union is exactly $\mathcal{S}$. Intuitively, each component represents a Reduce task, which is responsible for checking if any valid join result falls into it. Assume there are $k_R$ Reduce tasks, and the cardinality of relation $\mathcal{R}$ is denoted as $|\mathcal{R}|$. Then for each Reduce task, it has to check $\frac{\prod_{i=1}^{m}|\mathcal{R}_i|}{k_R}$ join results. However, it is not true that the more Reduce tasks, the less execution time. As when $k_R$ increases, the volume of data copy over network may grow significantly. For instance, as shown in Fig.5, when a Reduce task is added, the network volume increases.

Now we have the two sides of a coin, the number of Reduce tasks $k_R$ and partition function $f$. Our solution is described as follows. We first define what an "ideal" partition function is; then, we pick one such function and derive a proper $k_R$ for the given chain Theta-join query.

Let $t_{\mathcal{R}_i}^j$ denote the $j$-th tuple in relation $\mathcal{R}_i$. Partition function $f$ maps $\mathcal{S}$ to a set of $k_R$ components, denoted as $\mathcal{C}=\{c_1,c_2,...,c_{k_R}\}$. Let $Cnt(t_{\mathcal{R}_i}^j,\mathcal{C})$ denote the total number of times that $t_{\mathcal{R}_i}^j$ appears in all the components, we define the *partition score* of $f$ as

$$Score(f) = \sum_{i=1}^{n}\sum_{j=1}^{|\mathcal{R}_i|}Cnt(t_{\mathcal{R}_i}^j,\mathcal{C}) \quad (7)$$

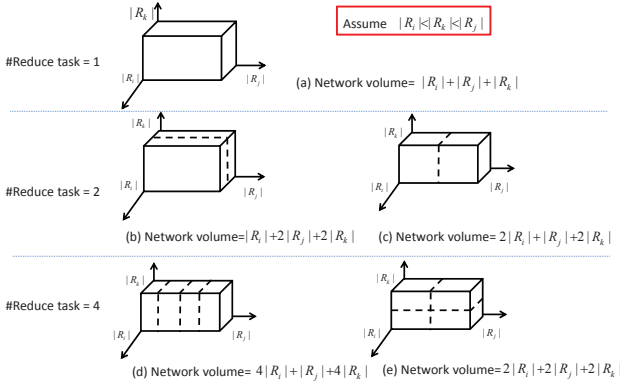

**Figure 5: How the network volume increases when more Reduce tasks are involved**

**Definition 5** *Perfect Partition Function. $f$ is a perfect partition function iff for a given $\mathcal{S}$, $\forall k_R$, $Score(f)$ is minimized.*

**Definition 6** *Perfect Partition Class. For a given $\mathcal{S}$, the class of all perfect partition functions, $\mathcal{F}$, is the perfect partition class of $\mathcal{S}$.*

Based on the definition of $\mathcal{F}$, to resolve $\mathcal{F}$ for a given $\mathcal{S}$ requires the "Calculus of Variation"[15], which is out of the scope of our current discussion. We shall directly present a partition function $f$ and prove that $f \in \mathcal{F}$.

**Theorem 2** *To partition a hyper-cube $\mathcal{S}$, the **Hilbert Space Filling Curve** is a perfect partition function $f$.*

PROOF. The minimum value of score function defined in Equ.7 is achieved when the following condition holds

$$\sum_{u=1}^{|\mathcal{R}_i|}Cnt(t_{\mathcal{R}_i}^u,\mathcal{C}) = \sum_{u=1}^{|\mathcal{R}_j|}Cnt(t_{\mathcal{R}_j}^u,\mathcal{C}) \ \forall 1 \le i,j \le n \quad (8)$$

In other words, in a partition component $c$, assume the number of distinct records from relation $\mathcal{R}_i$ is $c(\mathcal{R}_i)$, then the duplication factor of $\mathcal{R}_i$ in this component must be $\Pi_{j=1,j\neq i}^{n}c(\mathcal{R}_j)$. Since Hilbert space filling curve defines a traversing sequence of every cell in the hyper-cube of $\mathcal{R}_1 \times ...\mathcal{R}_n$, if we use a Hilbert curve $\mathcal{H}$ as a partition method, then a component $c$ is actually a continuous segment of $\mathcal{H}$. Considering the construction process of $\mathcal{H}$, every dimension is recursively divided by the factor of 2, and such recursive computation occurs the same number of times to all dimensions. Note that $\mathcal{H}$ defines a traversing sequence that traverses cells along each dimension fairly, meaning that if $\mathcal{H}$ has traversed half of $\mathcal{R}_i$, then $\mathcal{H}$ must also have traversed half of $\mathcal{R}_j$, where $\mathcal{R}_j$ is any other relation. Thus, given any partition value (equal to the number of Reduce tasks) $k_R$, then a segment of $\mathcal{H}$ of length $\frac{|\mathcal{H}|}{k_R}$, traverses the same proportion of records from each dimensions. Let this proportion be $\epsilon$. Therefore, the duplication factor for each record from $\mathcal{R}_i$ is

$$\Pi_{j=1,j\neq i}^{n}\eta\frac{|\mathcal{R}_j|}{2^\eta \times \epsilon} \quad (9)$$

where $\eta$ is the number of recursions. Note that the derived duplication factor satisfies the condition given in Equ.8. So $\mathcal{H}$ is a perfect partition function. $\square$

After obtaining $f$, we can further approximate the value of $k_R$ which achieves the best query evaluation time efficiency. As discussed earlier, $k_R$ affects two parts of our cost model, the network volume and the expected input size to Reduce tasks, both of which are the dominating factors of the execution time cost. Therefore, an approximation of the optimal $k_R$ can be obtained when we try to minimize the following value $\Delta$ (by computing the derivative of $k_R$). Notice that the first factor in Equ.10 is also a linear combination of $k_R$.

$$\Delta = \lambda \sum_{i=1}^{n}\sum_{j=1}^{|\mathcal{R}_i|}Cnt(t_{\mathcal{R}_i}^j,\mathcal{C}) + (1-\lambda)\frac{\prod_{i=1}^{m}|\mathcal{R}_i|}{k_R} \quad (10)$$

Intuitively, the $\Delta$ is a linear combination of the two cost factors. Coefficient $\lambda$ denotes the importance of each cost factor. For instance, if $\lambda < 0.5$, it implies that reducing the workload of each Reduce task brings more cost saving.[1] Note that the first cost factor in Equ.(10) is also a linear sum function of $k_R$. Therefore, by making $\Delta' = 0$, we can get $\lceil k_R \rceil$.

The pseudo code in Alg.1 describes our solution for evaluating a chain Theta-join query in one MRJ. Note that our main focus is the generation of (*key*,*value*) pairs. One tricky method we employed here, as also be employed in work [25], is randomly assigning an observed tuple $t_{\mathcal{R}_i}$ a global ID in $\mathcal{R}_i$. The reason is that, each Map task does not have a global view of the entire relation. Therefore, when a Map task reads a tuple, it cannot tell the exact position of this tuple in the relation.

---

**Algorithm 1:** Evaluating a chain Theta-join query in one MRJ

**Data:** Query $q = \mathcal{R}_1 \bowtie ... \bowtie \mathcal{R}_m$, $|\mathcal{R}_1|,...|\mathcal{R}_m|$;
**Result:** Query result
Using Hilbert Space Filling Curve to partition $\mathcal{S}$ and compute a proper value of $k_R$
Deciding the mapping: GlobalID($t_{\mathcal{R}_i}$)$\rightarrow$ a number of components in $\mathcal{C}$
**for** *each Map task* **do**
    GlobalID($t_{\mathcal{R}_i}$)$\leftarrow$ Unified random selection in $[1,|\mathcal{R}_i|]$
    **for** *all components that GlobalID($t_{\mathcal{R}_i}$) maps to* **do**
        generate (componentID, $t_{\mathcal{R}_i}$)

**for** *each Reduce task* **do**
    **for** *any combination of $t_{\mathcal{R}_1}$, ..., $t_{\mathcal{R}_m}$* **do**
        **if** *It is a valid result* **then**
            Output the result

---

### 5.2 Constructing $\mathcal{G}'_{JP}$

By applying the Alg.1, we can minimize the time cost to evaluate a chain Theta-join query using one MRJ. However, usually a group of MRJs is needed to evaluate multi-way Theta-joins. Therefore, we now discuss the construction of $\mathcal{G}'_{JP}$, which is a subgraph of the join-path graph $\mathcal{G}_{JP}$ and sufficient to serve the evaluation of N-join query $\mathcal{Q}$. As already discussed in Section 3.2, computing $\mathcal{G}_{JP}$ is a *#P-complete* problem, as it requires to enumerate all possible no-edge-repeating paths between any pair of vertices. In fact, only a subset of the entire edge collection in $\mathcal{G}_{JP}$ can be further employed in $\mathcal{T}_{opt}$. Therefore, we propose two pruning conditions to effectively reduce the search space in this section.

---
[1]In our experiments we observe that the value of $\lambda$ falls in the interval of (0.38,0.46). We set $\lambda$=0.4 as a constant.

The first intuition is that, to select $\mathcal{T}_{opt}$, the case that many join conditions are covered by multiple MRJs in $\mathcal{T}_{opt}$ is not preferred, because each join condition only needs to be evaluated once. However, it does not imply that MRJs in $\mathcal{T}_{opt}$ should strictly cover disjoint sets of join conditions. Because sometimes, by including extra join conditions, the output volume of intermediate results can be reduced. Therefore, we exclude a MRJ($e'_i$) on the only condition that there are other more efficient ways to evaluate all the join conditions that MRJ($e'_i$) covers. Formally, we state the pruning condition in Lemma 1.

**Lemma 1** *Edge $e'_i$ should not be considered if there exists a collection of edges ES, and the following conditions are satisfied: 1) $l'(e'_i) \subseteq \bigcup_{e'_j \in ES} l'(e'_j)$; 2) $w(e'_i) > Max_{e'_j \in ES} w(e'_j)$; 3) $s(e'_i) \geq \sum_{e'_j \in ES} s(e'_j)$.*

Lemma 1 is quite straightforward. If a MRJ can be substituted with some other MRJs that cover at least the same number of join conditions and be evaluated more efficiently with less demands on processing units, this MRJ cannot appear in $\mathcal{T}_{opt}$. Because $\mathcal{T}_{opt}$ is the optimal collection of MRJs to evaluate the query, containing any substitutable MRJ makes $\mathcal{T}_{opt}$ sub-optimal. For the second pruning method, we present the following Lemma which further reduces the search space.

**Lemma 2** *Given two edges $e'_i$ and $e'_j$, if $e'_i$ is not considered and $l'(e'_i) \subset l'(e'_j)$, then $e'_j$ should not be considered either.*

PROOF. Since $e'_i$ is not considered, it implies that there is a better solution to cover $l'(e'_i) \cap l'(e'_j)$. And this solution can be employed together with $l'(e'_j) - l'(e'_i)$, which is more efficient than computing $l'(e'_j)$ in one step. Therefore, $l'(e'_j)$ should not be considered. □

Note that Lemma 2 is orthogonal to Lemma 1. Since Lemma 1 decides whether a MRJ should be considered as a member of $\mathcal{T}_{opt}$, if the answer is negative, we can employ Lemma 2 to directly prune more undesired MRJs. By employing the two Lemmas proposed above, we develop an algorithm to construct $\mathcal{G}'_{JP}$ efficiently in an incremental manner, as presented in Alg.2.

---

**Algorithm 2:** Constructing $\mathcal{G}'_{JP}$

**Data**: $\mathcal{G}'_J$ containing $n$ vertices and $m$ edges, $\mathcal{G}'_{JP} = \emptyset$, a sorted list $WL = \emptyset$;
**Result**: $\mathcal{G}'_{JP}$
**for** $i=1:n$ **do**
  **for** $j > i$ **do**
    **for** $L=1:m$ **do**
4      **if** *there is a L-hop path from $\mathcal{R}_i$ to $\mathcal{R}_j$* **then**
        $e' \leftarrow$ the L-hop path from $\mathcal{R}_i$ to $\mathcal{R}_j$
      **if** $WL \neq \emptyset$ **then**
        scan $WL$ to find the first group of edges that cover $e'$
        apply Lemma 1 to decide if report $e'$ to $\mathcal{G}'_{JP}$
        **if** $e'$ *is not reported* **then**
          break //Lemma 2 plays the role
        insert $e'$ into $WL$ such that $WL$ maintains a sequence of edges in the ascending order of $w(e')$

---

Since we do not care the direction of a path, meaning $e'(v_i, v_j) = e'(v_j, v_i)$, we compute the pair-wise join paths following a fixed order of vertices (relations). In the Alg.2, we employ the linear scan of a sorted list to help decide whether a path should be reported in $\mathcal{G}'_{JP}$. One tricky part in the algorithm is line 4. A straightforward way is to employ DFS search from a given starting vertex, then the time complexity is $O(m+n)$. However, it introduces much redundant work for every vertex to perform this task. A better solution is before we run Alg.2, we firstly traverse $\mathcal{G}_J$ once and record the $L$-hop neighbor of every vertex. It takes only $O(m+n)$ time complexity. Then, line 4 can be determined in $O(1)$ time. Overall, we can see the worst time complexity of Alg.2 is $O(n^2 m)$. This happens only when $\mathcal{G}_J$ is a complete graph. In real practice, due to the sparsity of the graph, Alg.2 is quick enough to generate $\mathcal{G}'_{JP}$ for a given $\mathcal{G}_J$. As observed in the experiments, $\mathcal{G}'_{JP}$ can be generated in the time frame of hundreds of microseconds.

After $\mathcal{G}'_{JP}$ is obtained, we select $\mathcal{T}_{opt}$ following the methodology presented in [14], which gives $O(ln(n))$ approximation ratio of the optimum.

## 6. EXPERIMENTS

To verify the effectiveness of our solution, we conduct experiments on a real cluster environment with both real and synthetic data sets. In this section, we first describe the setup configuration of the test-bed and the data sets we used. Then we validate our cost model. We compare our solution for multi-way Theta-join processing with YSmart [23], Hive and Pig. We demonstrate that our solution can save on average 30% of query processing time when compared to the state of art methods. Especially in the cases of complex queries over huge volume of data, our method can save up to 150% of evaluation time.

### 6.1 Experiments Setup

Our experiments run exclusively on a 13-node cluster, where one node serves as the master node (Namenode). Every node has 2× Intel(R) Core(TM) i7 CPU 950 and 2× Kingston DDR-3 1333MHz 4GB of memory, 2.5TB HHD attached, running 2.6.35-22-server #35-Ubuntu SMP. All the nodes are connected with a 10GB-switch. In total, the test bed has 104 cores, 104GB main memory, and over 25TB storage capacity.

| Parameter Name | Default | Set |
|---|---|---|
| $fs.bloksize$ | 64MB | 64MB |
| $io.sort.mb$ | 100M | 512MB |
| $io.sort.record.percentage$ | 0.05 | 0.1 |
| $io.sort.spill.percentage$ | 0.8 | 0.9 |
| $io.sort.factor$ | 100 | 300 |
| $dfs.replication$ | 3 | 3 |

**Table 1: Hadoop parameter configuration**

We use Hadoop-0.20.205.0 to set up the system. Some major Hadoop parameters are given in Table 1, which follows the setting suggested by [21]. We use the *TestDFSIO* program to test the I/O performance of the system, and find that the system performance is stable, with average writing rate 14.69Mb/sec and reading rate 74.26Mb/sec. We run each experiment job 10 times and report the average execution time.

The first data set we employed for experiments is a real world data set collected from over 2000 mobile base stations from Oct 1st to Nov 30 in 2008. The data set records 571,687,536 phone calls from 2,113,968 different users. The data set contains 61 daily data files, which is of 20GB in total. The data schema is very simple, which contains the following five attribute fields:

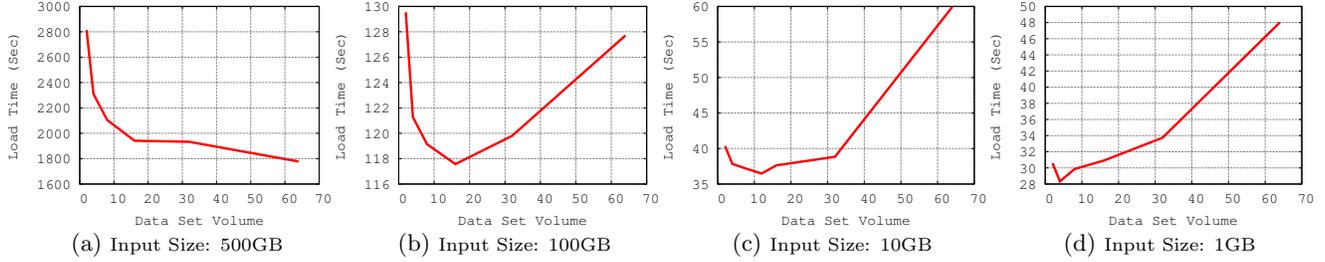

Figure 6: Execution time of sample *Join* task with different input size

| id | d: date | bt: begin time | l: length | bsc: base station code |

In the experiments, we design four queries of different complexities. We elaborate the workloads and complexity trend of the benchmark queries in Section 6.3.1. To validate the scalability of our solution, we enlarge the data set to 100GB and 500GB, by generating more phone calls, following the distribution of the number of phone calls along a day-time, which is a diurnal pattern (a periodical function with 24-hour cycles).

The second data set we employ is a synthetic but well recognized data set that specially designed for the TPC-H benchmark queries. We use the free tool DBGEN [1] to generate different size of testing data sets. We test almost all of the 21 benchmark queries that have multi-way join conditions. In this section we present the results of Q7, Q17, Q18 and Q21 to demonstrate the effectiveness of our solution, as they are well recognized benchmark queries to test how complex queries are evaluated. Since some queries only involve Equi-join, we slightly amend the join predicate to add inequality join conditions.

In the experiments, we compare our solution with YSmart, Hive and Pig. For the mobile data set, we develop the Hive and Pig scripts by ourselves. For the TPC-H test, we adopt the Hive codes from an open report[2], and develop efficient Pig scripts.

## 6.2 Cost Model Validation

The major factors affecting the performance of a MRJ are: 1) System parameter settings; 2) Input size and data set properties, especially the value distributions of join attributes; 3) Number of Reduce tasks. For the first factor, as elaborated in Section 4, we use random variable $p$ to denote the speed of spilling data to disk under different Map output ratio, and random variable $q$ to represent the cost of handling network connections of different number of Reduce tasks. We can predict the second factor by running data sampling algorithm (We conduct this task after data are uploaded to the HDFS). To decide $k_R$, a proper number of Reduce tasks, we can get a theoretical value of $k_R$ that guarantees the minimum execution time span by minimizing the cost formula (10). We validate the predication of $k_R$ with experiments and find that the optimal $k_R$ is mainly dominated by the output volume of Map tasks ($\lambda \approx 0.4$).

By studying the execution times of sample MRJs configured with different number of Reduce tasks, we get some insightful guidelines for selecting proper $k_R$. We run a sample MRJ conducting the join operation, which is included in Hadoop's standard release. We test different Map output sizes (1~200GB) and different value of $k_R$ (2~64). The results are shown in Fig.6. We find that, for a MRJ with large inputs, significant performance gains are obtained by increasing $k_R$ at the very beginning, as shown in Fig.6(a). However, when $k_R$ keeps growing up, performance gains become smaller and smaller. This phenomenon can be clearly observed from all four sub-figures in Fig.6. For a MRJ with relatively small input size, we see clear inflection point of performance when $k_R$ grows up, as observed in Fig.6(b), Fig.6(c) and Fig.6(d). Thus, we obtain a correlation between the input size (with Map output size determined) and $k_R$ for the best performance, as shown in Figure 7(a). We find that our experiment results can be well matched with a fitting curve (dashed line). We use this curve to determine $k_R$ for a given MRJ, such that we can compute the distribution of $p$ and $q$ which serve the estimation of a MRJ's running time. We compute $p$ and $q$ by studying an output controllable self-join program over a synthetic data set. Figure 7(b) gives the distributions of $p$ and $q$ according to different problem sizes.

[2]shttp://issues.apache.org/jira/secure/attachment/12416257/TPC-H on Hive 2009-08-11.pdf

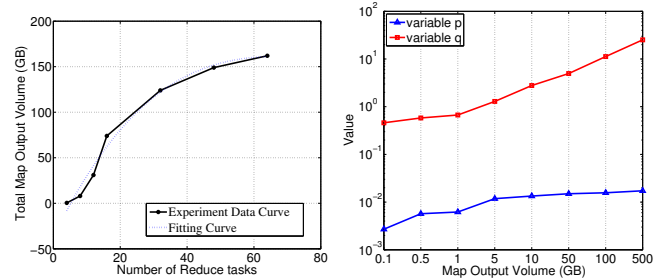

(a) Best $k_R$ for different size of Map ouput

(b) Value distributions of $p$, $q$

Figure 7: Selection of $k_R$, $p$ and $q$

To validate the effectiveness of our cost model, we checked the same self-join program over the mobile data set. As shown in the Fig.8, our estimation and the real MRJ execution time are very close.

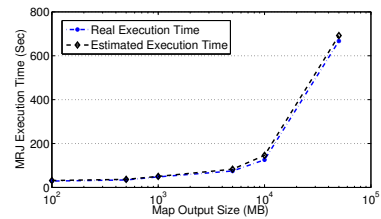

Figure 8: Cost model validation with a self-join program

## 6.3 Query Evaluation

Compare to YSmart and Hive, which targets at tables stored in Hive data warehouse, our solution targets at the plain data files stored in Hadoop Distributed File System (HDFS). In addition to simply upload the data to HDFS, we run a sampling algorithm to collect rough data statistics and build the index structure, which is the reason that our method is a little more time consuming for the data uploading process, as shown in Fig.11. For comparison, we also present the cost for simply uploading data files to HDFS. Note that the uploading is performed by each DataNode from their local disk. Comparing with Hive, our method demonstrates comparable time cost of data uploading for large data volumes.

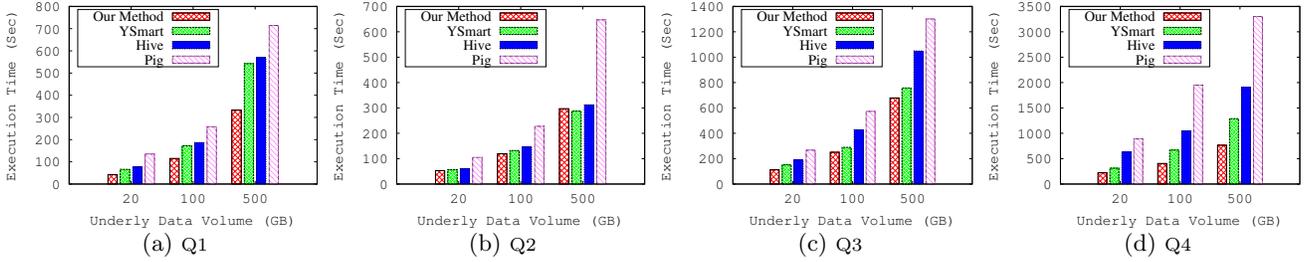

Figure 9: Execution time of 4 queries over the mobile data set in different scales, $k_P \leq 96$

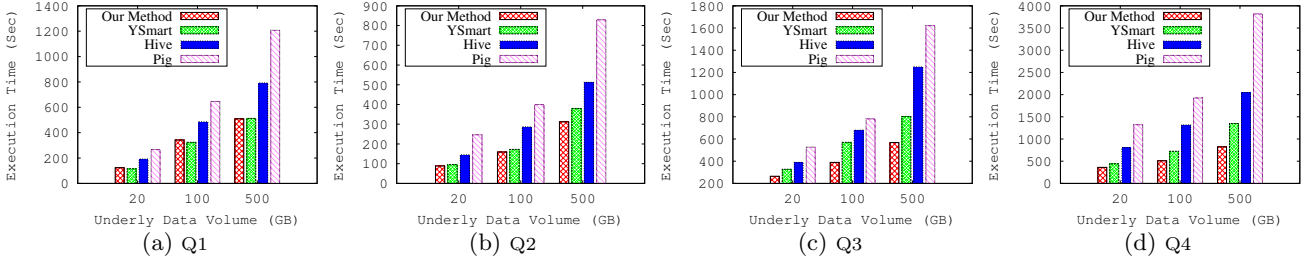

Figure 10: Execution time of 4 queries over the mobile data set in different scales, $k_P \leq 64$

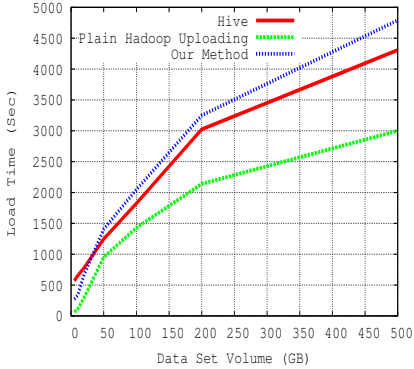

Figure 11: The time cost for data loading

### 6.3.1 Real World Mobile Data

We design four multi-way Theta-join queries for the mobile data set, which are of different complexities in terms of covering different inequality functions and joining on attributes with different selectivities. We test benchmark queries on different scales of data volumes to validate the scalability of our solution. In comparison, we also test YSmart, Hive and Pig scripts that perform the same tasks.

For comprehensiveness, we describe the four queries in a SQL-like style as follows:

**Q1** SELECT $t_3.id$ FROM table $t_1$, table $t_2$, table $t_3$ WHERE $t_1.bt \leq t_2.bt$, $t_1.l \geq t_2.l$, $t_2.bsc=t_3.bsc$, $t_2.d=t_3.d$

**Q2** SELECT $t_3.id$ FROM table $t_1$, table $t_2$, table $t_3$ WHERE $t_1.bt \leq t_2.bt$, $t_1.l \geq t_2.l$, $t_2.bsc \neq t_3.bsc$, $t_2.d=t_3.d$

**Q3** SELECT $t_1.id$ FROM table $t_1$, table $t_2$, table $t_3$, table $t_4$, WHERE $t_1.d < t_2.dt$, $t_2.d < t_3.dt$, $t_1.d+3 > t_3.d$, $t_1.bsc=t_4.bsc$

**Q4** SELECT $t_1.id$ FROM table $t_1$, table $t_2$, table $t_3$, table $t_4$, WHERE $t_1.d < t_2.dt$, $t_2.d < t_3.dt$, $t_1.d+3 > t_3.d$, $t_1.bsc \neq t_4.bsc$

In plain English, the first two queries return the concurrent phone calls for the same base station and all concurrent phone calls at different base stations, respectively. The third query returns the user whose calls are handled by the same base station 3-day in a row. The fourth query finds out the user whose calls are handled by different base stations 3-day in a row. These queries can help monitor the workload distribution among base stations and capture unusual behavior of customers. Table 2 summarizes the features of the benchmark queries. Note that the four queries are listed in the ascending order of running time complexity. As shown in the table, the benchmark queries being employed cover all the inequality functions and have significant differences in output size.

| $Q$ | Relations Cnt. | Inequality Func. | Join Cnt. | Result Sel. |
|---|---|---|---|---|
| $Q1$ | All | $\{\leq, \geq\}$ | 3 | 0.00035 |
| $Q2$ | All | $\{\leq, \geq, \neq\}$ | 3 | 0.00108 |
| $Q3$ | All | $\{<, >\}$ | 4 | 0.00079 |
| $Q4$ | All | $\{<, >, \neq\}$ | 4 | 0.01524 |

Table 2: Benchmark query statistics

As we elaborate in Section 3, there may not be enough processing units to evaluate queries in the most time-saving fashion. Therefore, we test the benchmark queries by specifying different number of available processing units, as shown in Fig.9 and Fig.10, respectively. The results shown in Fig.9 demonstrate that our solution has comparable time cost comparing with the state of art method YSmart. Especially when the query is relatively easy, like Q1 and Q2, our solution at best gives near YSmart performance. The reason lies in two folds. First, for simple queries, there is little optimization opportunity for MRJ scheduling. Second, YSmart take multiple inter-MRJ optimization techniques into consideration, which is not the focus of our work. In this case, compare to Hive and Pig, the time saving of our solution lies in eliminating unnecessary network volumes and redundant Reduce task workloads.

When we specify $k_P$ (the number of available processing units) to be at most 64, the advantage of our solution for more complex queries are obvious. As shown in Fig.10, take

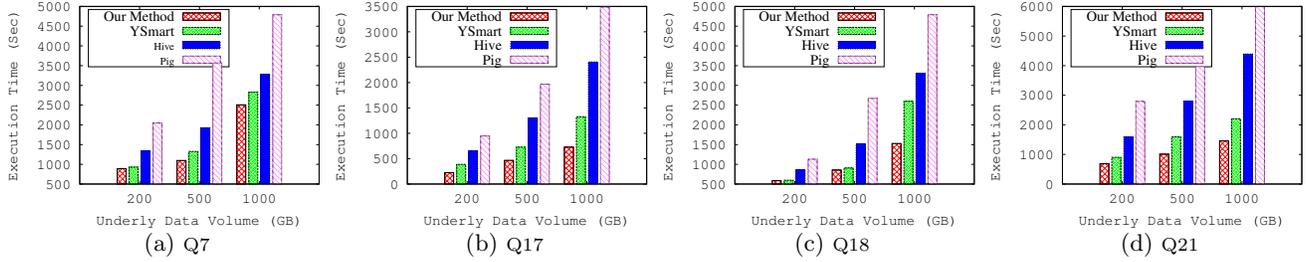

Figure 12: Execution time of 4 TPC-H benchmark queries in different scales, $k_P \leq 96$

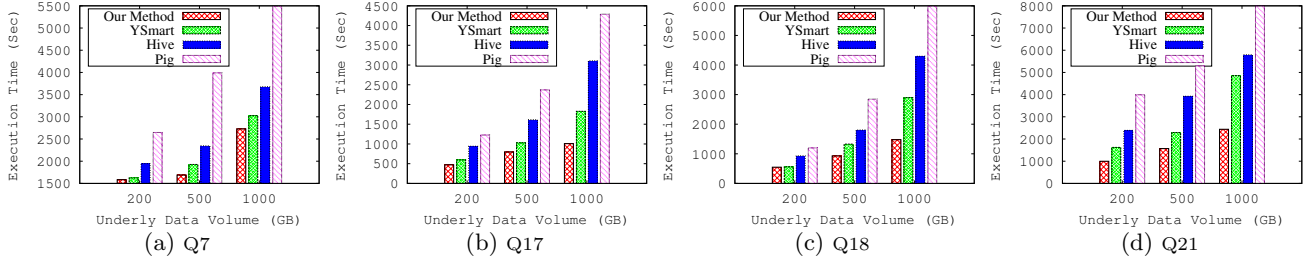

Figure 13: Execution time of 4 TPC-H benchmark queries in different scales, $k_P \leq 64$

Q4 for instance, our solution achieves about 50% time savings comparing to YSmart.

### 6.3.2 TPC-H Benchmark Queries

We test almost all 21 benchmark queries from the TPC-H benchmark, excluding some simple queries that only involve two or three relations and simply join on foreign keys, like Q1 and Q2. In this section we present the result of 4 queries, which are well recognized complex queries for performance test. In the experiments, we also run the query under different available number of processing units. The results are presented in Fig.12 and Fig.13. Table 3 summarizes the features of the 4 benchmark queries.

| $\mathcal{Q}$ | Relations Cnt. | Inequality Func. | Join Cnt. | Result Sel. |
|---|---|---|---|---|
| $Q7$ | 5 | $\{\leq, \geq\}$ | 8 | 0.00176 |
| $Q17$ | 3 | $\{\leq\}$ | 4 | 0.00426 |
| $Q18$ | 4 | $\{\geq\}$ | 4 | 0.00021 |
| $Q21$ | 6 | $\{\geq, \neq\}$ | 8 | 0.00087 |

**Table 3: TPC-H query statistics**

When we consider all processing units are involved in the evaluation, as reported by Fig.12, we have the following observations. First, as reported in [23], YSmart generally has over 200% speedup comparing to Hive. Second, by taking the advantage of index structures and data statistics, our solution for the multi-way Theta-join queries have 30% of time savings on average compare to YSmart. The reason is that, our solution try to minimize the data copying volume over network and balance the workload of Reduce tasks. Third, for the case that the number of process units is sufficient, i.e., when the involved data volume is relatively small, our method gains more time saving by taking the advantage of the "greedy" scheduling, as shown in Fig.12(b) and Fig.12(c). Moreover, along with the increasing of data set volume, our solution also demonstrates satisfactory scalability as Hive does.

When we set $k_P$ to a smaller value, e.g. $\leq 64$, Fig.13 shows that our method achieves even more time saving comparing to a larger $k_P$ ($k_P \leq 96$). For instance, as shown in Fig.13(a) and Fig.13(d), along with the growth of underlying data volumes, our method demonstrates better scalability. Since our solution employs $k_P$-aware scheduling of MRJs, when $k_P$ is changed, the selection of $\mathcal{T}$ and execution plan are updated correspondingly. On the contrary, Hive always try to employ as many Reduce tasks as possible to perform a task, and YSmart does not take this factor into consideration. Therefore, we observe as much as 150% speedup comparing to the YSmart solution.

In summary, as expected and proved by experiments, our solution wins the state of art solutions in two aspects: 1) when there is not enough processing units, our solution is able to dynamically choose a near optimal solution to minimize the evaluation makespan; 2) Our solution takes the advantages of data statistics and index structures to guide the (*key,value*) partition among Reduce tasks. On one hand, we eliminate unnecessary data copying to perform a Theta-join query. On the other hand, we minimize the redundant computation in Reduce tasks. Therefore, in the context of fitting multi-way Theta-join evaluation in a dynamic Cloud computing platform, our solution demonstrates promising scalability and execution efficiency.

## 7. RELATED WORK

Existing efforts toward efficient join query evaluation using MapReduce mainly fall into two categories. The first category is to implement different types of join queries by exploring the partition of (*key, value*) pairs from Map tasks to Reduce tasks without touching the implementation details of the MapReduce framework. The second category is to improve the functionality and efficiency of MapReduce itself to achieve better query evaluation performance. For example, MapReduce Online [9] allows pipelined job interconnections to avoid intermediate result materialization. A PACT model [4] extends the MapReduce concept for complex relational operations. Our work, as well as work [27] on set similarity join, work [25] on Theta-join, all fall in the first category. We briefly survey some most related works in this category.

F.N.Afrati and at el. [2] present their novel solution for evaluating multi-way Equi-join in one MRJ. The essential idea is that, for each join key, they logically partition the Reduce tasks into different groups such that a valid join re-sult can be discovered on at least one Reduce task. Their optimization goal is to minimize the volume of data copying over the network. But the solution only works for the Equi-join scenario. Because for Equi-join, as long as we make the join attribute the partition key, the *joinable* data records that have the same key value will be delivered to the same Reduce task. However, for Theta-join queries, such partition method for (*key,value*) pairs cannot even guarantee the correctness. Moreover, answering complex join queries with one MRJ may not guarantee the best time efficiency in practice. Wu Sai and et al. [28] targets at the efficient processing of multi-way join queries over massive volume of data. Although they present their work in the context of Equi-join, their focus is how to decompose a complex query to multiple MRJs and schedule them to eventually evaluate the query as fast as possible. However, their decomposition is still join-key oriented. Therefore, after decomposing the original query into multiple pair-wise joins, how to select the optimal join order is the main problem. On the contrary, although we also explore the scheduling of MRJs in this work, each MRJ being scheduled can involve multiple relations and multiple join conditions. Therefore, our solution truly tries to explore all possible evaluation plans. Moreover, work [28] does not take the limit of processing unit into consideration, which is a critical issue in real practice. Some other works try to explore the general work flow of single MRJ or multiple MRJs to improve the whole throughput performance. Hadoop++ [13] injects optimized UDFs into Hadoop to improve query execution performance. RCFile [17] provides a column-wise data storage structure to improve I/O performance in MapReduce-based warehouse systems. MRShare [24] explores the optimization opportunities to share the file scan and partition key distribution among multiple correlated MRJs. YSmart [23] is a source-to-source SQL to MapReduce translator. It proposes a common-MapReduce framework to reduce redundant file I/O and duplicated computation among Reduce tasks. Recent system works presented query optimization and data organization solutions that can avoid high-cost data re-partitioning when executing a complex query plan, like SCOPE [29] and ES$^2$ [7].

## 8. CONCLUSION

In this paper, we focus on the efficient evaluation of multi-way Theta-join queries using MapReduce. Our solution includes two parts. First, we study how to conduct a chain-type multi-way Theta-join using one MapReduce job. We propose a Hilbert curve based space partition method that minimizes data copying volume over network and balances the workload among Reduce tasks. Second, we propose a resource aware scheduling schema that helps the evaluation of complex join queries achieves a near optimal time efficiency in resource restricted scenarios. Through extensive experiments over both synthetic and real world data, our solution demonstrates promising query evaluation efficiency comparing to the state-of-art solutions.

## 9. ACKNOWLEDGMENTS

This work is supported in part by the Hong Kong RGC GRF Project No.611411, HP IRP Project 2011, National Grand Fundamental Research 973 Program of China under Grant 2012-CB316200, Microsoft Research Asia Grant, MRA11EG05 and HKUST RPC Grant RPC10EG13.